\documentclass[letterpaper, 10 pt, conference]{ieeeconf}  
\IEEEoverridecommandlockouts             
\usepackage{amsmath, amssymb, amsfonts}
\usepackage[colorlinks=true, citecolor=blue, linkcolor=blue]{hyperref}
\let\labelindent\relax
\usepackage{url, footnote, enumitem}
\usepackage{tikz}\usetikzlibrary{calc, positioning}
\usepackage{graphicx, xcolor}
\usepackage{caption, subcaption}

\usepackage{amsmath, amsfonts, amssymb, amsthm}
\allowdisplaybreaks
\theoremstyle{plain}
\newtheorem{remark}{Remark}

\newtheorem{definition}{Definition}

\newtheorem{theorem}{Theorem}

\title{\LARGE \bf Learning the Integral Quadratic Constraints on Plant-Model Mismatch}
\author{Wentao Tang
\thanks{This work is supported by the National Science Foundation (Award \#2414369).}
\thanks{Wentao Tang is an assistant professor with the Department of Chemical and Biomolecular Engineering, North Carolina State University, Raleigh, NC 27695, U.S.A. {\tt\small wentao\_tang@ncsu.edu}}
}
\begin{document}
\maketitle\thispagestyle{empty}\pagestyle{empty}

\begin{abstract}
    While a characterization of \emph{plant-model mismatch} is necessary for robust control, the mismatch usually can not be described accurately due to the lack of knowledge about the plant model or the complexity of nonlinear plants. 
    Hence, this paper considers this problem in a data-driven way, where the mismatch is captured by parametric forms of \emph{integral quadratic constraints} (IQCs) and the parameters contained in the IQC equalities are learned from sampled trajectories from the plant. 
    To this end, a one-class support vector machine (OC-SVM) formulation is proposed, and its generalization performance is analyzed based on the statistical learning theory. 
    The proposed approach is demonstrated by a single-input-single-output time delay mismatch and a nonlinear two-phase reactor with a linear nominal model, showing accurate recovery of frequency-domain uncertainties. 
\end{abstract}

\section{\textsc{Introduction}}
\par Model-based control, with model predictive control \cite{rawlings2018model} as a representative scheme, is known to be the mainstream control strategy in practice, especially for multivariable constrained control. Due to the inherent complexity of plant dynamics, the existence of disturbances and noises, as well as the expense of system identification, plant models are often far from precise, which leaves the identification of \emph{plant-model mismatch} a major long-lasting issue \cite{van1998closed, badwe2010quantifying}. The problem of controller performance monitoring (or mismatch \emph{detection}) \cite{qin1998control, qin2007recent, huang2008dynamic, kaw2014improved, gao2016review} is closely related to the mismatch identification -- typically, when the controller performance is found to be largely deteriorated, then mismatch needs to be characterized and compensated in the nominal model. Here, we focus on the identification problem. 

\par From a pragmatic point of view, most model-based control schemes applied in industrial practice are \emph{linear model-based} (see, e.g., \cite{prett1987shell, tang2023automatic}), where transfer functions are used to represent input-output relations. Thus, we desire that plant-model mismatch is characterized in terms of the error on the nominal models in a Laplace domain. 
Indeed, if the plant and the nominal model can be both considered as transfer functions, then the mismatch can be identified using informative input signals with rich perturbations in the frequency range of interest \cite{badwe2009detection, ling2017detection}. 
However, the actual plant dynamics may not be a linear one; arguably, if the plant was almost linear, then the identification of the plant model should have been accurate, making the mismatch detection and identification of less value. Thus, the plant-model mismatch is better described as one \emph{between a nonlinear plant and a linear nominal model}. This can be addressed via information-theoretic \cite{chen2020framework}, Gaussian process \cite{wu2021online}, or deep learning \cite{son2022learning, moayedi2024physics} approaches. 
However, a nonlinearly described mismatch (e.g., a neural network) can be difficult to utilize in linear control, requiring significant changes or even completely replacement of the control algorithm. 

\par Hence, we aim to characterize the underlying nonlinear plant-model mismatch ``in a linear way'', so as to allow the controller design or tuning algorithm to remain in a linear framework. This is conceptually related to the classical problem of specifying the conditions for nonlinear systems to be robustly controlled by linear controllers. 
There exist a range of tools for this purpose, from specific to generic -- absolute stability, Popov criteria, passivity, dissipativity, and integral quadratic constraint (IQC) \cite{sastry1999nonlinear}. 
Simply speaking, these conditions serve as ``bounds'' on the nonlinear non-idealities, giving rise to linear/quadratic inequalities to constrain the uncertainties and thus enabling the synthesis of robust control laws as linear ones \cite{zhou1998essentials, dullerud2001course, lozano2013dissipative}. 
In particular, the most generic characterization mentioned here, the IQC of a system, refers to the existence of dynamic multipliers on its input and output signals, such that the signal from the dynamic multipliers satisfy a quadratic dissipative inequality \cite{megretski1997system, seiler2014stability, veenman2016robust}. 
Hence, the problem of interest is \emph{how to learn the IQC on the plant-model mismatch from process input and output data}. 

\par The problem that we encounter here is similar to the ones pertaining to the learning of $L_2$-gain, passivity index, and dissipativity of an unknown (black-box) nonlinear system. For linear systems, these learning problems were explored based on Willems' Fundamental Lemma \cite{wahlberg2010non, berberich2020trajectory, koch2021provably}. In the author's previous works \cite{tang2019input, tang2019dissipativity, tang2021dissipativity}, machine learning techniques are proposed for learning the dissipativity of nonlinear systems. A recent work from Bridgeman and her coworkers \cite{locicero2024issues} gave preliminary analysis of the statistical learning theory underpinning dissipativity learning. 

\par In this paper, we adopt the one-class support vector machine (OC-SVM) approach \cite{tang2019input} for IQC learning and provide a  generalization performance bound (similar to \cite{locicero2024issues}). Specifically, in the IQC, the dynamic multiplier is fixed by choosing basis transfer functions (filters), so that the inequality specifying the IQC is parameterized by a symmetric matrix with a positive definite input diagonal block and a negative definite output diagonal block. 
The IQC is then expressed as linear inequality constraints involving such parameters and sampled trajectories of the plant as data. The learning of the parameters through OC-SVM yields the desired IQC characterization of the plant-model mismatch on the frequency domain. 

\par The remaining paper is organized as follows. Preliminaries on nonlinear control theory is provided in \S\ref{sec:preliminaries}. The proposed technique is then discussed in \S\ref{sec:learning}. A simple numerical example and a practical application are shown in \S\ref{sec:example} and \S\ref{sec:application}, respectively\footnote{Codes at available at the author's GitHub repository (\url{https://github.com/WentaoTang-Pack/IQClearning}). 
}. Conclusions are given in \S\ref{sec:conclusion}. 

\par \textit{Notations.} Upper case letters are used to represent matrices, transfer functions, or dynamical systems, and lower case letters are for scalars and column vectors. 
For a matrix $A\in \mathbb{R}^{n\times n}$ whose entries are written as $a_{ij}$ ($1\leq i, j\leq n$), its trace is $\mathrm{tr}\,A = \sum_{i=1}^n a_{ii}$ and its Frobenius norm is $\|A\|_\mathrm{F} = \left[\sum_{i=1}^n\sum_{j=1}^n |a_{ij}|^2\right]^{1/2}$. The inner product between two $n\times n$ matrices $A$ and $B$ is $\langle A, B\rangle:= \mathrm{tr}(A^\top B)= \sum_{i=1}^n\sum_{j=1}^n a_{ij}b_{ij}$. We denote by $A\succeq B$ for two symmetric matrices if $A-B$ is positive semidefinite. $I$ represents the unit matrix of appropriate dimension, or a static system where the outputs is identical to the inputs.
For a complex scalar (or vector, or matrix) $a$ ($A$), we denote by $a^\dagger$ ($A^\dagger$) its conjugate (or conjugate transpose). We use $j=\sqrt{-1}$.

\section{\textsc{Preliminaries}}\label{sec:preliminaries}
\par We consider an \emph{unknown} plant dynamics (in the scope of this paper, the plant-model mismatch as a system itself) in the form of a nonlinear continuous-time system
$$\Sigma: \begin{cases}
\dot{x}(t) = f(x(t), u(t)) \\
y(t) = h(x(t), u(t))
\end{cases} $$
defined on $t\in [0, +\infty)$, where $x(t)\in\mathbb{R}^{n_x}$, $u(t)\in\mathbb{R}^{n_u}$, and $y(t)\in\mathbb{R}^{n_y}$ are the states, inputs, and outputs, respectively. $f$ and $h$ are Lipschitz. We denote the Laplacian transforms of the time-domain signals by replacing $t$ with $s$ without changing the function symbol, e.g., $u(s) = \int_0^\infty u(t)e^{-st}dt$. Consider a matrix of stable proper real rational transfer functions $\Psi \in \mathcal{RH}_\infty^{n_z\times (n_y+n_u)}$, called a \emph{dynamic multiplier}, that act on inputs and outputs separately: 
$$ z(s) = \Psi(s)\begin{bmatrix} y(s) \\ u(s) \end{bmatrix} = \begin{bmatrix} \Psi_y(s) & 0 \\ 0 & \Psi_u(s) \end{bmatrix} \begin{bmatrix} y(s) \\ u(s) \end{bmatrix} = \begin{bmatrix}
    z_y(s) \\ z_u(s)
\end{bmatrix}. $$
% \footnote{If a state-space realization of $\Psi$ can be found as $(A^\Psi, B^\Psi, C^\Psi, D^\Psi) = (A^\Psi, [B^\Psi_y, B^\Psi_u], C^\Psi, [D^\Psi_y, D^\Psi_u])$, i.e., $\Psi(s) = C^\Psi(sI-A^\Psi)^{-1}B^\Psi + D^\Psi$, then the aggregated dynamics of the plant $\Sigma$ and the dynamic multiplier $\Psi$ as illustrated in Fig. \ref{fig:dynamics} can be written as:
% \begin{align*}
% (\Sigma, \Pi): \begin{cases}
% \dot{x}(t) = f(x(t), u(t))\\
% y(t) = h(x(t), u(t)), \\
% \dot{\xi}(t) = A^\Psi \xi(t) + B^\Psi_y y(t) + B^\Psi_u u(t) \\ 
% z(t) = C^\Psi\xi + D^\Psi_y y(t) + D^\Psi_u u(t) 
% \end{cases}
% \end{align*}
% has $z$ as its outputs.} 
\begin{figure}
	\centering
	\begin{tikzpicture}
	\draw[thick, red!75, fill=red!20] (-0.5, -0.5) rectangle (1.0, 0.5); 
	\node[red] at (0.25, 0) {$\Sigma$}; 
	\draw[thick, black, -latex] (-2, 0) -- (-0.5, 0);
	\node[black] at (-2.25, 0) {$u$};
	\draw[thick, black, -latex] (1.0, 0) -- (2.25, 0);	\node[black] at (2.5, 0) {$y$};
	\draw[thick, blue!75, fill=blue!20] (2, -2) rectangle (3.5, -1);
	\node[blue] at (2.75, -1.5) {$\Psi(s)$}; 
	\draw[thick, black, -latex] (-1, 0.0) -- (-1, -1.75) -- (2, -1.75); 
	\draw[thick, black, -latex] (1.5, 0) -- (1.5, -1.25) -- (2, -1.25);
	\draw[thick, black, -latex] (3.5, -1.5) -- (4.75, -1.5);
	\node[black] at (5, -1.5) {$z$};
	\end{tikzpicture}
	\caption{The plant and dynamic multiplier}\label{fig:dynamics}
    \vspace{-1.0em}
\end{figure}
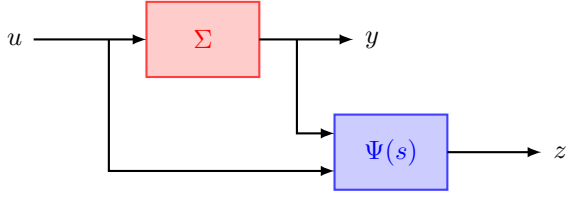
\begin{definition}\label{def:dissipativity}
    The system $\Sigma$ is said to be \emph{dissipative under the dynamic multiplier} $\Psi: (y, u)\mapsto z$ with respect to the supply rate $\sigma(z) = z^\top Mz$, if for any trajectory of the aggregated dynamics $(\Sigma, \Psi)$ starting from the origin ($x(t_0)=0$, $\xi(t_0)=0$) and for any $t_1\geq t_0$,  we have 
    \begin{equation}\label{eq:dissipativity}
        \int_{t_0}^{t_1} z^\top(t) M z(t)dt \geq 0
    \end{equation}
    where $M$ is a symmetric real matrix. 
\end{definition}

\par When all the components of $u(\cdot)$, $y(\cdot)$ and $z(\cdot)$ are $L_2$-signals (square integrable on $[0,+\infty)$), a necessary condition for \eqref{eq:dissipativity} is that it holds when $t_1-t_0\rightarrow+\infty$. According to the Parseval's identity, this implies the following inequality, which involves a corresponding quadratic form on frequency domain integrated throughout the imaginary axis, called an \emph{integral quadratic constraint} (IQC):
\begin{equation}\label{eq:IQC}
\int_{-\infty}^{+\infty} \begin{bmatrix} y^\dagger(j\omega) & u^\dagger(j\omega) \end{bmatrix} \Pi(j\omega) \begin{bmatrix} y(j\omega) \\ u(j\omega) \end{bmatrix} \geq 0,
\end{equation}
where $\Pi(s) = \Psi^\dagger(s)M\Psi(s)$. However, \eqref{eq:dissipativity} is defined for any $T\geq 0$. Hence, \eqref{eq:dissipativity} is a stronger condition than \eqref{eq:IQC}. Therefore, \eqref{eq:dissipativity} is also referred to as the \emph{hard IQC} and \eqref{eq:IQC} is called the corresponding soft IQC. $(\Psi, M)$ is said to be a \emph{hard factorization} of the IQC specified by $\Pi$ \cite{seiler2014stability}. 
\begin{remark}
    When the supply rate function takes a direct quadratic form of $(y, u)$, i.e., using a ``trivial'' dynamic multiplier $\Psi = I$: 
    $$\sigma(y, u) = \begin{bmatrix} y^\top & u^\top \end{bmatrix} M \begin{bmatrix} y \\ u \end{bmatrix}$$
    for some symmetric matrix $M$, we say that the system is $M$-dissipative. If $M = \begin{bmatrix} Q & S\\ S^\top & R \end{bmatrix}$, the system is $(Q, S, R)$-dissipative. In particular, the system is passive if $n_u=n_y$, $Q = R=0$, and $S = I$. 
\end{remark}
\begin{remark}
    The transfer function entries in $\Psi_y(s)$ and $\Psi_u(s)$ can be viewed as operators for \emph{feature extraction} from signals. For example, for a SISO system, if $\Psi_y=1$, $\Psi_u(s)=1/(1+\tau s)$ ($\tau>0$), $M_{yy}=M_{uu}=0$, $M_{yu}=M_{uy}=1/2$, then $\Sigma$ can be interpreted as a \emph{passive} system with respect to the output and a first-order filtered input instead of the original input. 
\end{remark}

\par Similar to the approach of \cite{hill1976stability}, one can establish the following conclusion that an IQC system conforming to Definition \ref{def:dissipativity} (i.e., one dissipative under a dynamic multiplier $\Psi$) should possess a storage function of the internal states. The proof is identical to the one in \cite{hill1976stability} or \cite{tang2019dissipativity}, except that $\Sigma$ should now be trivially substituted with $(\Sigma, \Psi)$. 
\begin{theorem}\label{thm:Willems}
    If $\Sigma$ is dissipative under the dynamic multiplier $\Psi$ with respect to the supply rate $\sigma(z)$ as specified in Definition \ref{def:dissipativity}, then there exists a positive semidefinite function $V(x, \xi)$, defined for any $(x, \xi)$ that is reachable from the origin through some input trajectories $u(\cdot)$ and satisfying $V(0, 0) = 0$, such that the dissipative inequality: 
    $$ V(x(t_1), \xi(t_1)) - V(x(t_0), \xi(t_0)) \leq \int_{t_0}^{t_1} \sigma(z(t)) dt$$
    holds for any trajectory on any time interval $[t_0, t_1]$. Such a function $V$ is called the storage function, and can be constructed according to
    $$ V(x, \xi) = \inf_{\substack{(u(t),d(t)), \enskip t\in[0,T] \\ x(0)=0, \enskip x(T) = x, \enskip \xi(0)=0, \enskip \xi(T) = \xi}} \int_0^T \sigma(z(t))dt. $$
\end{theorem}

\par The knowledge of IQC on the uncertainty of a system facilitates the analysis of robust stability, robust performance, and the design of the desired robust controllers (see, e.g., \cite{megretski1997system, veenman2014iqc}). % We shall not elaborate on the review of these topics in the present paper. Interested readers are referred to, e.g., the works of Veenman and Scherer \cite{veenman2014iqc, scherer2022dissipativity}, Seiler \cite{seiler2014stability}, and Khong \cite{khong2021integral}. 
It should, however, be pointed out that without a full model (or even \emph{with} a nonlinear model), it is generally difficult to determine the IQC. Thus, in the context of identifying plant-model mismatch, we consider the problem of \emph{learning (estimating, inferring) the IQC from data}.

\section{\textsc{Learning of IQC Parameters}}\label{sec:learning}
\subsection{Problem Setting}
\par Consider a plant $\Pi$, as an input-output map ($u \mapsto y$) whose dynamics is unknown and in general nonlinear. Its nominal plant is denoted as $\Pi_0$, for which the output under input $u$ is denoted as $y_0$. The plant-model mismatch $\Delta = \Pi - \Pi_0$ has an output $r=y-y_0$, called the \emph{residual}. 
More generally, we may also consider multiplicative mismatch, i.e., $\Delta$ such that $\Pi = (1+\Delta)\Pi_0$, namely $y-y_0 = \Delta y_0$, or other types of mismatch. Hereforth in this section, we denote the input and output of $\Delta$ as $v$ and $w$, respectively. We assume that the mismatch $\Delta: v\mapsto w$ satisfies some IQC specified by $(\Psi, M)$. 

\par For simplicity, we may assume that the dynamic multiplier $\Psi(s)$, comprising of feature extraction operators, are given and fixed, and thus only the symmetric matrix $M$ is to be estimated. 
Specifically, some filters are adopted in the construction of $\Psi_w$ and $\Psi_v$ to transform each component of $w$ and $v$:
\begin{equation}\label{eq:dynamic.multiplier}
    z(s) = \Psi(s)\begin{bmatrix} w(s) \\ v(s) \end{bmatrix} = \begin{bmatrix} \Psi_w(s) & 0 \\ 0 & \Psi_v(s) \end{bmatrix} \begin{bmatrix} w(s) \\ v(s) \end{bmatrix} = \begin{bmatrix}
    z_w(s) \\ z_v(s)
\end{bmatrix}. 
\end{equation}

\begin{remark}[Choice of filters]
    Without additional prior information, we may use a finite number of M{\"{u}}ntz-Laguerre filters: $\varphi_1(s) = \frac{\sqrt{2\mathrm{Re}\,b_1}}{s+b_1}$, $\varphi_k(s) = \frac{\sqrt{2\mathrm{Re}\,b_k}}{s+b_k} \prod_{q^\prime=1}^{k-1} \frac{s-\bar{b}_{k^\prime}}{s+b_{k^\prime}}$ ($k\geq 2$) with preassigned poles $-b_1, -b_2, \dots$ whose real parts do not exceed $-\epsilon$ for some $\epsilon>0$ and satisfying $\sum_{k=1}^\infty \frac{\mathrm{Re}\, b_k}{1+|b_k|^2} = \infty$. They are known to form a uniformly bounded orthonormal basis of the Hardy space $\mathcal{H}_2$ \cite{knockaert2001orthonormal}. 
    One may also consider the use of a combination of low-pass, high-pass, and band-pass filters to extract inputs on different frequency ranges. In the choice of filters, the assignment of poles are expected to be critical for accuracy. 
\end{remark} 

\begin{definition}
    The matrix $M$ is called the \emph{dissipativity parameters}. % The range of valid dissipativity parameters for the system is called the \emph{dissipativity set} and denoted as $\mathcal{M}$. 
\end{definition}
For the estimation of the dissipativity parameters $M$, we suppose that $m$ trajectories $(v^{(i)}(\cdot), w^{(i)}(\cdot))_{i=1}^m$ are sampled independently, from a distribution of signals with random time durations $t_1-t_0$ and input excitations. We assume that all such trajectories start from the origin. 
Formally, we denote this distribution as a measure $\mathbb{P}$. The goal is therefore to determine a valid choice of $M$ such that for all $i=1,\dots,m$, the following inequality holds approximately: 
$$\int_{t_0^{(i)}}^{t_1^{(i)}} z^{(i)\top}(t) Mz^{(i)}(t)dt \geq 0. $$

\subsection{OC-SVM for IQC Learning}
Rewriting the inequality \eqref{eq:dissipativity} as 
$$ \bigg\langle M, \int_0^T z(t)z^\top(t) dt \bigg\rangle = \mathrm{tr}\left(M\int_0^T z(t)z^\top(t)dt\right) \geq 0, $$
the goal is to find $M$ such that the above inequality approximately holds on the sampled trajectories. 
\begin{definition}
    For a trajectory $(v(\cdot), w(\cdot))$, which determines a trajectory of $z(\cdot)$ on $[0, T]$ according to \eqref{eq:dynamic.multiplier}, its corresponding dual dissipativity parameters refers to
    \begin{equation}\label{eq:dual.dissipativity.parameters}
	\Gamma = \int_0^T z(t)z^\top(t) dt \enskip (\succeq 0).
    \end{equation}
\end{definition}
Hence, having calculated the dual dissipativity parameters of the sampled trajectories $\{\Gamma^{(i)}\}_{i=1}^m$, we seek $M$ such that $\langle M, \Gamma^{(i)}\rangle \geq 0$ approximately. 
This problem is amenable to one-class support vector machine (OC-SVM) \cite{scholkopf2001estimating}, where we maximize the ``margin'' of the inequality, i.e., a nonnegative value $\rho>0$ such that $\langle M, \Gamma_i\rangle \geq \rho$ for all $i$, but penalizing the norm of the ``slope'', i.e., $\|M\|_\mathrm{F}$. Equivalently, the problem is to minimize $\|M\|_\mathrm{F}^2$ while rewarding $\rho$. 
For a more flexible formulation, we allow the margin $\rho$ to be violated, while the violations by each sampled trajectory are to be penalized as a cost. In the following so-called ``soft-SVM'' formulation, $\nu \in (0, 1)$ is the ``softness'' constant and $\xi_i\geq 0$ ($1\leq i\leq m$) are the margin violations. 
\begin{equation}\label{eq:OC-SVM}
    \begin{aligned}
        \min_{M, \rho, \xi} \, & \frac{1}{2}\|M\|_\mathrm{F}^2 -\rho + \frac{1}{\nu m}\sum_{i=1}^m \xi_i \\
        \mathrm{s.t.} \, & \langle M, \Gamma_i \rangle \geq \rho - \xi_i,  \, \xi_i \geq 0\, (1\leq i\leq m); \\
        & M = \begin{bmatrix}
            M_{ww} & 0 \\ 0 & M_{vv} 
        \end{bmatrix}, \, -M_{ww} \succeq \epsilon_wI, \, M_{vv} \succeq \epsilon_vI. 
    \end{aligned}
\end{equation}

\par Here, instead of using the typical SVM formulation, we should impose additional constraints on the blocks of $M$ for our IQC learning problem. The last line of \eqref{eq:OC-SVM} guarantees that (i) the ``output'' of the mismatch system $\Delta$ (i.e., the output residual $r$) contributes to a decrease in the storage function, so that the mismatch is a self-stabilized system\footnote{If $M_{ww}$ turns out to be a scalar, then the negative definiteness constraint can be simplified as $M_{ww}=-1$. Similarly, if $M_{vv}$ is scalar, then only $M_{vv}=1$ is needed.}, (ii) the inputs can only lead to an increase in the storage, and that (iii) the absence of input-output bilinear terms in the supply rate for simplicity (in fact, they can always be relaxed by an arithmetic-geometric mean inequality). 

\begin{remark}[Soft OC-SVM]
    The use of a soft OC-SVM that allows the nonnegative margin to be violated is justified by the following two considerations. 
    First, we assumed that all such trajectories start from the origin (equilibrium point), which is dificult to guarantee or verify. Practically, the sampled trajectories have nonzero initial values in the storage function, which can cause violation of the dissipative inequality. 
    Second, the system is assumed to be noiseless and disturbance-free, while actual plants are always perturbed. 
\end{remark}

\begin{remark}[Sampling the trajectories]
    The generation of informative sample trajectories is critical for learning the IQC. In \cite{tang2021dissipativity} for dissipativity learning, it was proposed that the input excitations are created by randomly sampling the Fourier coefficients, and the trajectory duration is selected such that the output ranges over an interval of interest. 
    In \cite{locicero2024issues}, a variety of sampling methods, including the Fourier coefficient \cite{tang2021dissipativity}, Legendre polynomial, and Wiener process, are experimented for passivity index learning. % indicating that the appropriate choice of time duration and a sufficient large dictionary of basis functions are desired. 
    Since IQC is a frequency-domain characterization, in this work, it is recommended that the input signal $v(t)$ be sampled to cover the frequency range of interest. An explicit approach is to let $v$ be sinusoidal waves of randomized frequencies, which is to be used in \S\ref{sec:example} and \S\ref{sec:application}. 
\end{remark}

\subsection{Generalization Performance}
\par The probabilistic guarantee on the generalization error of OC-SVM was given in Sch{\"o}lkopf et al. \cite{scholkopf2001estimating}. 
\begin{theorem}
    Suppose that the OC-SVM gives an optimal solution $(M^\ast, \rho^\ast, \xi_i)$ with $\rho_\ast = \rho^\ast - \xi_i^\ast > 0$ for all $1\leq i\leq m$. Then for any $\delta \in (0, 1)$ and $\epsilon>0$, with probability $1-\delta$ (over the choice of samples of size $m$), we have
    $$\mathbb{P}\left[\langle M, \Gamma\rangle < \rho_\ast - \epsilon\right] \leq \frac{2}{m}\left(\lceil \log \kappa(\|M\|_\mathrm{F}, \,\epsilon) \rceil  + \frac{2}{\delta}\right)$$
    where $\kappa(\alpha, \epsilon, 2m)$ is the covering number of the model space $\{\langle M, \cdot\rangle: \|M\|_\mathrm{F}\leq \alpha\}$ by balls of radius of $\epsilon$ under the metric $d(M_1, M_2) = \sup_{\Gamma_1, \dots, \Gamma_{2m}} \max_{i=1,\dots, 2m} |\langle M_1-M_2, \Gamma_i\rangle|$. 
\end{theorem}
In LoCicero \cite{locicero2024issues}, the conclusion was translated explicitly in terms of the dissipativity learning that has the same form as our IQC learning problem formulated in \eqref{eq:OC-SVM}. Specifically, by the definition of covering number, it could be found that 
$$\mathbb{P}\left[\langle M, \Gamma\rangle < \rho_\ast - \epsilon\right] \leq \frac{1}{m} \mathcal{O}\left(\log \frac{1}{\delta} + \frac{1}{\epsilon^2}\log m + \frac{1}{\epsilon^2} \log \frac{1}{\epsilon}\right). $$

\section{\textsc{Numerical Example}}\label{sec:example}
\par Consider a simple case where the actual SISO system has an unknown delay not accounted for in the nominal model. The delay is $\theta\in[0, \theta_0]$ where the upper bound $\theta_0$ is known. As pointed out in \cite{megretski1997system}, the plant-model mismatch $\Delta(s)=e^{-\theta s}-1$ (as a multiplicative uncertainty) satisfies IQCs of the form
\begin{equation}\label{eq:IQC.structure}
    \Pi(j\omega)=\begin{bmatrix}
    -\tau(j\omega) & 0 \\ 0 & \tau(j\omega)\ell(j\omega)
\end{bmatrix}
\end{equation}
where $\tau(j\omega)\geq 0$ is a rational weighting function and $\ell$ is a real-valued rational function of $\omega$ satisfying $\ell(j\omega) \geq \ell_0(j\omega)$, in which
$$\ell_0(j\omega) = \max_{\theta\in[0, \theta_0]} |e^{-j\omega\theta}-1|^2 = \begin{cases}
    4\sin^2 \frac{\omega\theta_0}{2}, & \omega< \pi/\theta_0 \\
    4, & \omega\geq \pi/\theta_0
\end{cases}.$$
Such majorization does exist, e.g., $\ell(\omega)=(4\omega^4 + 50\omega^2)/(\omega^4 + 6.5\omega^2 + 50)$ in Megretski and Rantzer \cite{megretski1997system}. 
For simplicity, let $\tau\equiv1$ be fixed and $\psi$ be learned from data. It is of interest to examine whether the learned dynamic multiplier is indeed of the theoretical form above. 

\par Without loss of generality, say $\theta_0 = 1/2$. 
To generate the data, we let $u(t)$ be a sinusoidal wave $u(t)=A\sin \omega t$. Here we choose $A=1$ and $\log_{10}\omega$ is sampled from a uniform distribution on $[-2, 2]$, in order to cover a sufficiently large range of frequencies. For IQC learning, $m=500$ trajectories are collected. 
We choose three transfer functions: $\varphi_1(s)=1/(s+1)$ (low-pass filter), $\varphi_2(s)=s/(s+1)$ (high-pass filter), and $\varphi_3(s)=\varphi_1(s)\varphi_2(s)$ (band-pass filter) to extract the input features, i.e., let $z_1=y$ and $z_{k+1}=\varphi_k(s)u$ ($k=1,2,3$). We aim to obtain a matrix of the form $M = \mathrm{diag}(-1, M_u)$ where $M_u$ is a positive semidefinite $3\times 3$ matrix. 

\begin{figure}[!t]
    \centering
    \includegraphics[width=\columnwidth]{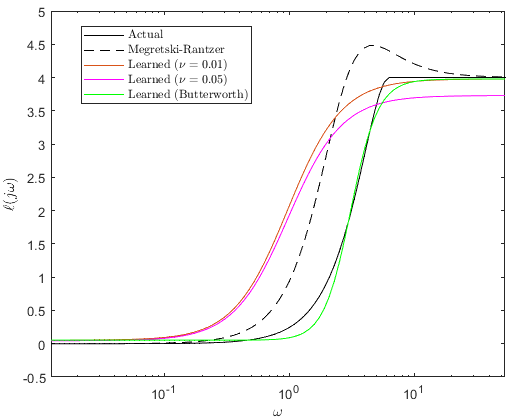}
    \caption{Frequency in the learned IQC of the example system with a delay mismatch.}
    \label{fig:delay_learned}
    \vspace{-1.5em}
\end{figure}
\par By solving the resulting optimization problem \eqref{eq:OC-SVM} in \texttt{cvx} (version 2.2 in Matlab R2024b), the $M$ matrix is found as 
$$M = \begin{bmatrix}
    0.0245 & -0.0120 & 0.0221 \\
   -0.0120 & 3.9616 & 0.0158 \\
    0.0221 & 0.0158 & 0.0227 \\
\end{bmatrix} \approx \mathrm{diag}(0, 4, 0)$$
which gives $\ell(j\omega)\approx 4\omega^2/(1+\omega^2)$. The comparison of the learned IQC is compared with the theoretical one as well as the lowest possible curve $\ell_0(j\omega)$ in Fig. \ref{fig:delay_learned}. 
Here the softness parameter $\nu=0.01$ is chosen, which results in a small violation of the OC-SVM margin $\sum_{i=1}^m \xi_i/m=9.34\times10^{-4}$. When $\nu$ is increased to $0.05$, the average violation becomes $3.07\times10^{-2}$, causing more high-frequency learning error. This indicates that OC-SVM is a suitable learning tool, at least when the dataset is clean and informative. 

\par From the result, it is found that the learned IQC provides an approximately correct characterization of the underlying uncertainty, in the sense that except when $\omega\gtrsim\pi/\theta_0$, the IQC learned is an overestimation, and also upon $\omega\rightarrow0$ and $\omega\rightarrow\infty$, the limits are close to the actual values. 
On the other hand, since the learned IQC dominantly relies on the high-pass features of the input to provide the S-shaped curve, the pole that is assigned to $\varphi_2(s)$, which is $1$, misaligns with the half-rise frequency $\pi/2\theta_0$ of $\ell_0(j\omega)$. Also, the rise of $\ell_0$ is steeper than a first-order high-pass filter. 
Hence, to attempt for a tighter estimation, we set $\varphi_1$ as the second Butterworth filter with cutoff frequency $\pi/2\theta_0$, and $\varphi_2$ as its high-pass counterpart. The resulting $\ell$ is shown in Fig. \ref{fig:delay_learned} as well. 
As expected, the learned IQC becomes more accurate; however, this assumes prior knowledge on a better pole assignment and better filter choice. 

\par Empirically, we found that the learning result is insensitive to the sampling strategy in this simple example. When sampling $u(t)$ as random binary sequences (with a discretization time of $0.05$) or as truncated Fourier series with $5$ sinusoidal waves of uniformly distributed coefficients, the IQC learned remain identical to the ones shown above. 
Such robustness should be due to the \textit{a priori} selection of a correct IQC structure \eqref{eq:IQC.structure} that relies on the user's judgment.

\section{\textsc{Application to a Chemical Process}}\label{sec:application}
\begin{figure}[!t]
     \centering
     \begin{subfigure}[b]{0.45\columnwidth}
         \centering
         \includegraphics[width=\textwidth]{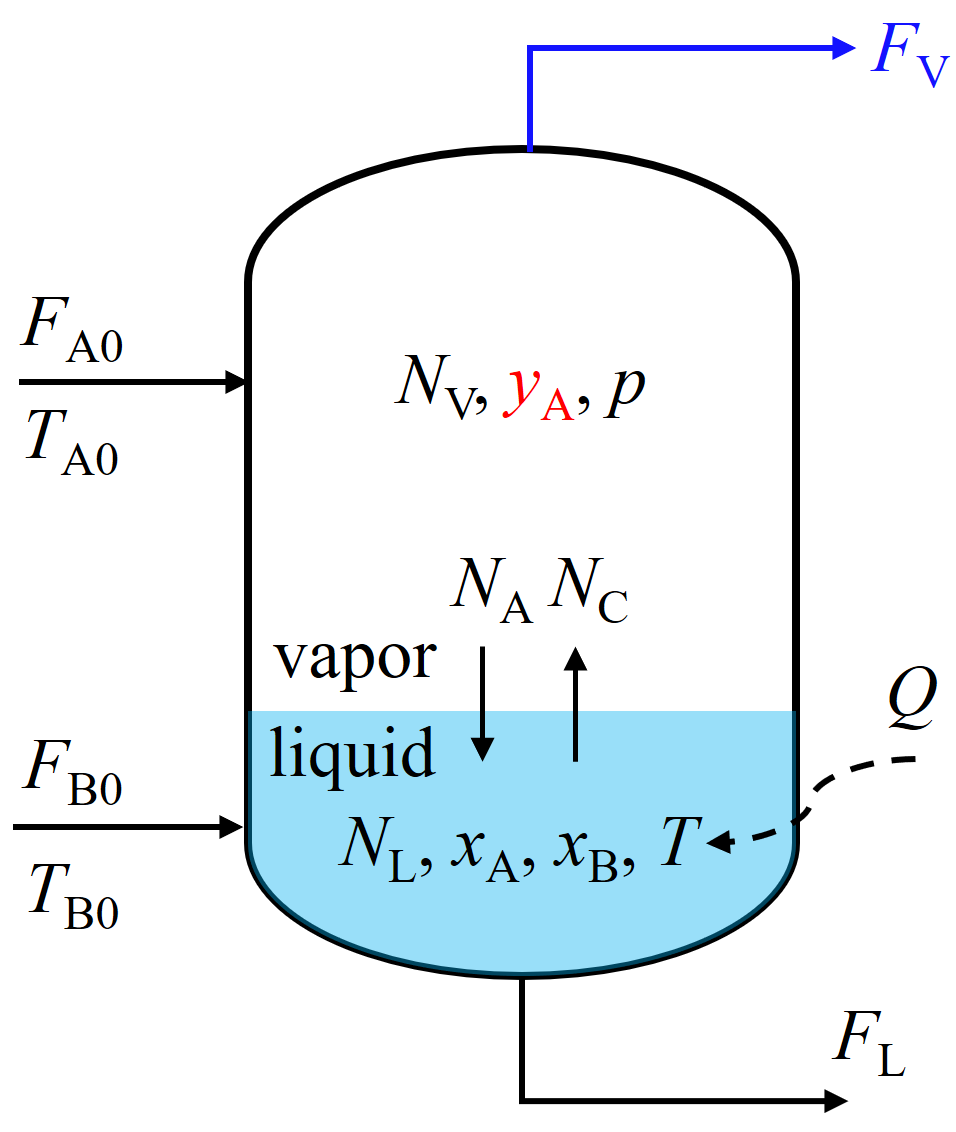}
         \caption{Process schematic}\label{fig:twophase}
     \end{subfigure}
     \\
     \begin{subfigure}[b]{\columnwidth}
         \centering
         \includegraphics[width=\textwidth]{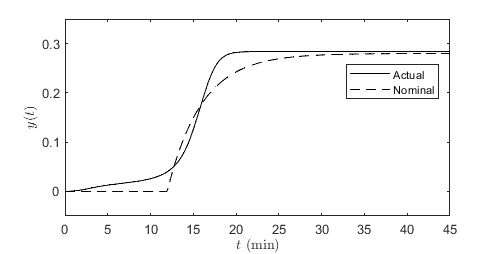}
         \caption{Unit step response}\label{fig:twophase_step}
     \end{subfigure}
     \caption{The two-phase reactor.}
     \vspace{-1.5em}
\end{figure}
\par Consider the two-phase reactor in \cite{kumar1995feedback}. An illustration of the process is provided in Fig. \ref{fig:twophase} and the underlying true model, which is nonlinear, was detailed in \cite{tang2021dissipativity}. We focus on the relation between the vapor flow rate $F_\text{V}$ as an input ($u$) and the substrate concentration in the vapor phase $y_\text{A}$ as an output ($y$). The step response of this process is shown in Fig. \ref{fig:twophase_step}. 
Suppose that from this step response, a simplistic engineer considers the delayed first-order transfer function $\Pi_0(s)=K_0e^{-\theta_0s}/(\tau_0s+1)$ as the linear nominal model, where $K_0=0.28$, $\tau_0 = 4$, and $\theta_0=12$. The step response of the nominal model is plotted in contrast to the actual step response in Fig. \ref{fig:twophase_step}. We are thus interested in characterizing the nonlinear plant-model mismatch $\Delta = \Pi - \Pi_0$. 

\par We sample $m=500$ trajectories from the unknown nonlinear dynamics under input excitations $u(t) = A\sin\omega t$ for a duration of $45$ min, where $\log_{10} \omega \tau_0$ comes from a uniform distribution on $[-2, 2]$ and $A=1/4$. Under these settings, the simulations are numerically stable. 
To learn the IQC, we choose the IQC structure as in \eqref{eq:IQC.structure} with $\tau(j\omega)\equiv 1$ and $M = \mathrm{diag}(1, M_u)$. Thus, $\ell(j\omega) = \Psi_u(j\omega)^\dagger M_u \Psi_u(j\omega)$. 
The filters in $\Psi_u(s)$ are decided in the following way: (i) $9$ frequencies ($\omega_1=10^{-1}, \omega_2=10^{-3/4}, \cdots, \omega_9=10^1$) are first chosen; (ii) between each two frequencies $\omega_k$ and $\omega_{k+1}$, let $\varphi_k(s) = \frac{s/\omega_k}{s/\omega_k + 1}\cdot \frac{1}{s/\omega_{k+1}+1}$; and (iii) let $\varphi_0(s) = \frac{1}{s/\omega_1+1}$ and $\varphi_{10}(s) = \frac{s/\omega_9}{s/\omega_9 + 1}$. 

\par
\begin{figure}[!t]
    \centering
    \includegraphics[width=\columnwidth]{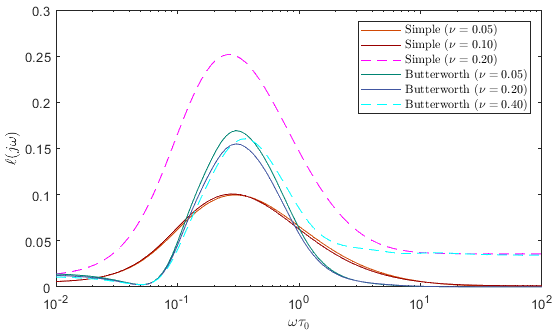}
    \caption{Frequency response $\ell(j\omega)$ in the learned IQC for the two-phase reactor.}
    \label{fig:twophase_learned}
    \vspace{-1.5em}
\end{figure}

 The resulting $\ell(\omega)$ of the learned IQC under different SVM softness parameters $\nu$, as well as when the high-pass and low-pass filters are substituted by the Butterworth second-order ones, are shown in Fig. \ref{fig:twophase_learned}. 
Similar to as observed in the previous section, when using simple filters, the curve of $\ell(j\omega)$ tend to be less steep, while Butterworth second-order filters better concentrate the frequency-domain uncertainties. 
With large $\nu$, the violation to the linear inequality constraints in the OC-SVM problem \eqref{eq:OC-SVM} can cause an erroneous high-frequency mismatch identification. It is therefore necessary to adopt a small enough $\nu$ to recover the anticipated conclusion that the mismatch should not be significant at very low or very high frequencies. 

\par
\begin{figure}[!t]
    \centering
    \includegraphics[width=\columnwidth]{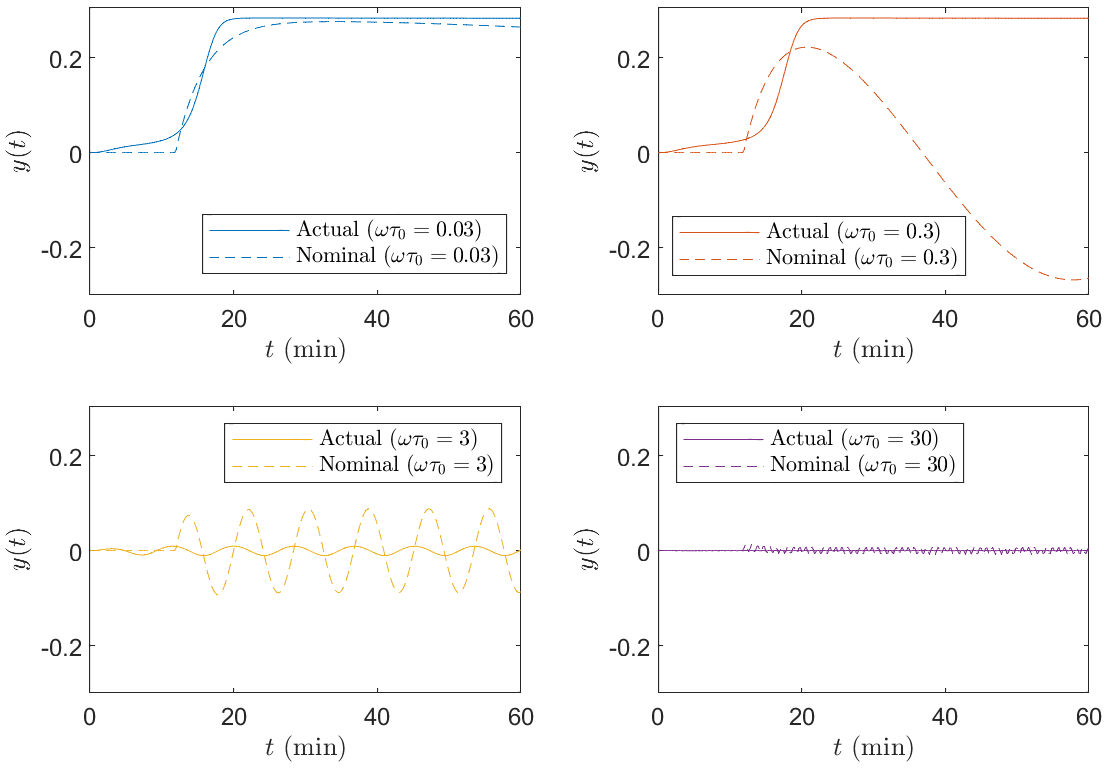}
    \caption{Comparison of actual and nominal responses under different frequencies.}
    \label{fig:twophase_verification}
    \vspace{-1.5em}
\end{figure}

 It thus appears from Fig. \ref{fig:twophase_learned} that at a frequency $\omega\tau_0\approx 0.3$, the mismatch peaks. For a verification that the mismatch is indeed most severe around this frequency, we compare the actual and nominal responses under $u(t)=\cos\omega t$ for  $\omega\tau_0 = 0.03, 0.3, 3, 30$, shown in Fig. \ref{fig:twophase_verification}. 
One can intuitively observe here that \emph{while the nominal response is a delayed wave, the actual nonlinear dynamics does not even exhibit any oscillation} for $\omega\tau_0=0.3$. Therefore, we conclude that the proposed IQC learning approach indeed provides an accurate description of the plant-model mismatch on the frequency domain. 

\section{\textsc{Conclusions}}\label{sec:conclusion}
In this work, a practical and efficient-to-implement approach to characterize the mismatch between a nonlinear plant and its nominal model in the form of a parameterized IQC. A numerical example and a realistic chemical process application demonstrate its capacity to accurately recover the mismatch information on the frequency domain. 
The IQC learned allows the synthesis of robust model-based controllers. With increasing practice of machine learning in the context of data-driven control \cite{tang2022data}, it is envisioned that the proposed method can find its use in state-of-the-art control technology. For nonlinear MPC, possible extensions of the current approach to corresponding nonlinear model structures will be studied in the future research. 

% \section*{APPENDIX}
\section*{\textsc{Acknowledgment}}
The author thanks Dr. Pierre Carrette, Advanced Process Control R\&D Lead at Shell Global Solutions, whom the author worked for several years ago, for the exposure to plant-model mismatch detection and identification problems. 

\bibliographystyle{ieeetr}
\bibliography{mybib.bib}
\end{document}